%% file: main.tex
\documentclass[preprint,1p,numbers,sort&compress]{elsarticle}
\usepackage{amssymb}
\usepackage[linktocpage=true]{hyperref}
\usepackage{mathtools}
\usepackage{slashed}
\usepackage{url}
\usepackage{graphicx}
\usepackage{cleveref}
\usepackage[compat=1.0.0]{tikz-feynman}
\usepackage{tikz-feynman}
\usepackage{overpic}
\usepackage{subfigure}
\usepackage{diagbox}
\usepackage{multirow}

\usepackage{siunitx}
\input{units}

\journal{Astroparticle Physics}

\begin{document}

\begin{frontmatter}
    \title{Atmospheric pion, kaon, and muon fluxes for sub-orbital experiments}
    \author[iowa]{Diksha Garg\corref{cor1}}\ead{diksha-garg@uiowa.edu}
    \author[iowa]{Laksha~Pradip~Das}
    \author[iowa]{Mary~Hall~Reno}
    \affiliation[iowa]{
    organization={University of Iowa},
    addressline={30 N Dubuque St},
    city={Iowa City},
    state={IA},
    postcode={52242},
    country={United States of America}
    }
    \cortext[cor1]{Corresponding author}

    \begin{abstract}
        Cosmic rays interacting with the Earth's atmosphere generate extensive air showers, which produce Cherenkov, fluorescence and radio emissions. These emissions are key signatures for detection by ground-based, sub-orbital, and satellite-based telescopes aiming to study high energy cosmic ray and neutrino events. However, detectors operating at ground and balloon altitudes are also exposed to a background of atmospheric charged particles, primarily pions, kaons, and muons, that can mimic or obscure the signals from astrophysical sources.
        In this work, we use coupled cascade equations to calculate the atmospheric pion, kaon and muon fluxes reaching detectors at various altitudes. 
        Our analysis focuses on energies above 10 GeV, where the influence of the Earth's magnetic field on particle trajectories is minimal.
        We provide angular and energy-resolved flux estimates and discuss their relevance as background for extensive air shower detection. Our results are potentially relevant for interpreting data from current and future balloon-borne experiments such as EUSO-SPB2 and for refining trigger and veto strategies in Cherenkov and fluorescence telescopes.
    \end{abstract}
\end{frontmatter}

\tableofcontents

\section{Introduction}
We are currently in the era of multimessenger astronomy, where cosmic messengers such as cosmic rays, gamma rays, gravitational waves, and neutrinos are studied in combination to uncover the nature and origin of their astrophysical sources~\cite{Meszaros:2019xej}. Over the past decade, balloon-borne experiments like Extreme Universe Space Observatory on a Super Pressure Balloon 1 (EUSO-SPB1)~\cite{JEM-EUSO:2023ypf} and 2 (EUSO-SPB2)~\cite{SPB2,Eser:2021mbp,Adams:2025owi} and upcoming POEMMA Balloon with Radio (PBR)~\cite{ICRC2025Eser} have emerged as important tools for studying very-high-energy (VHE) cosmic rays and neutrinos ($E > 10^{15}$ eV$=10^6$ GeV). These sub-orbital detectors, deployed at 33 km altitude, are designed to observe extensive air showers (EASs) that can be initiated either by cosmic ray interactions in the atmosphere or from neutrinos interacting within the Earth to produce charged leptons exiting the Earth to decay in the atmosphere. EASs can travel upward toward the balloon-based instruments and emit signatures such as optical Cherenkov light, radio signals, and fluorescence radiation. Such emissions can be detected by sub-orbital, and orbital observatories including ANITA (I-IV)~\cite{ANITA:2008mzi,ANITA:2010hzc,ANITA:2018sgj,ANITA:2019wyx}, PUEO~\cite{PUEO:2020bnn}, EUSO-SPB1, EUSO-SPB2, PBR and the proposed POEMMA~\cite{Olinto_2021} and M-EUSO~\cite{ICRC2025Plebaniak} missions. The Trinity ground-based imaging air Cherenkov telescope~\cite{Bagheri:2025fxh} also targets skimming neutrinos that produce upward-going EASs with low angles relative to the Earth.

\begin{figure*}[t]
    \centering
    \includegraphics[width=0.45\textwidth]{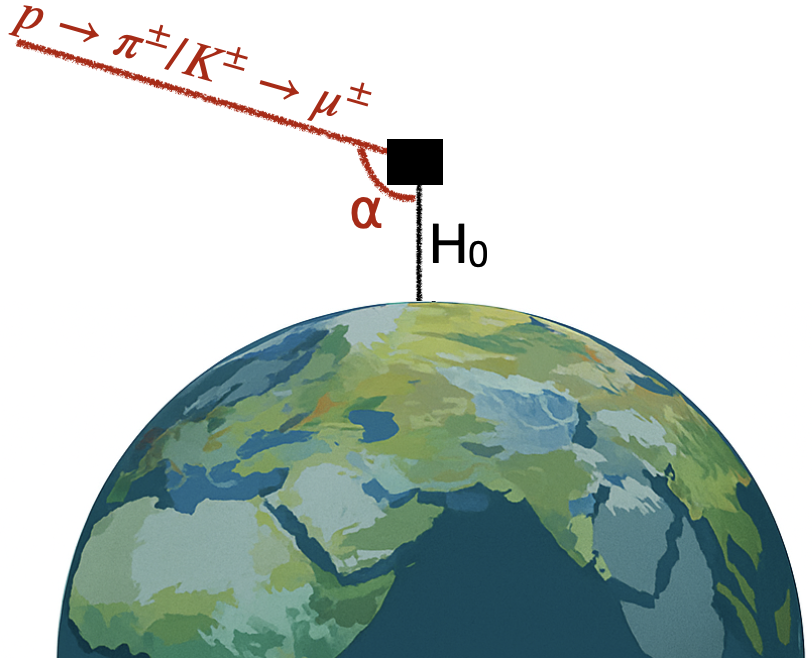}
    \caption{Detector (shown as black box) at altitude H$_{0}$ (not to scale). 
    In red is the particle trajectory to the detector's surface which makes angle $\alpha$ as shown in the figure ($\alpha=180^\circ$ for down-going particles incident on the detector). The cosmic ray nucleons (e.g., $p$) interact in the atmosphere to create pions and kaons which may decay to muons. 
    }
    \label{fig:Earth_model}
\end{figure*}

An additional by-product of cosmic ray interactions in the atmosphere is the generation of atmospheric muons and neutrinos~\cite{Gaisser:2002jj}, produced primarily via decays of charged mesons. Charged pions and kaons are most relevant for energies of interest here.   The charged mesons and atmospheric muons can directly strike detectors on-board balloon experiments and also on the ground, 
potentially mimicking or contaminating EAS signals. The telescope-Earth geometry is illustrated in~\cref{fig:Earth_model} for an instrument at altitude $H_0$, where $\alpha$ represents the incident cosmic ray angle relative to the telescope. 
Here, $\alpha=180^\circ$ means the cosmic ray is vertically incident (down-going), $\alpha=90^\circ$ for horizontally incident cosmic rays, and $\alpha<90^\circ$ for upward-going cosmic rays. 
For $H_0=18,\ 33$ km, hadrons and muons produced in the atmosphere can be incident only with $\alpha>85.7^\circ,\ 84.2^\circ$, respectively, since cosmic rays do not transit through the Earth. While atmospheric lepton fluxes have been extensively studied for ground-based detectors~\cite{Lipari:1993hd,Gaisser:2019xlw,Gaisser:2002jj,Fedynitch:2015zma,Fedynitch:2018cbl,Kozynets:2023tsv}, there has been less attention devoted to their impact on sub-orbital instruments. In this paper, we present an estimate of the atmospheric muon and charged meson flux directly hitting the sub-orbital telescopes at $H_0=33$ km. This is a potentially relevant background to electromagnetic signals of cosmic ray-induced EASs. We  compare fluxes at $H_0=33$ km with fluxes reaching the ground and also at $H_0=18$ km of altitude where EUSO-SPB2 flight took data on its second day of flight while it was losing altitude as the mission was terminated~\cite{Adams:2025owi}. This work is an extension of the preliminary results shown in ref.~\cite{Garg:2023gay}.

In~\cref{sec:2}, we outline the overall strategy and tools used to evaluate particle fluxes in this work. In~\cref{sec:results}, we present our results on how the fluxes of charged pions, kaons, and muons vary with detector altitude. We also examine the ratio of pion to kaon contributions, both to the muon flux and to their own respective total fluxes reaching the detector. The section concludes with calculations of the atmospheric fluxes for different telescope orientations, followed by a discussion on the main sources of uncertainties affecting the meson and muon flux predictions. Finally, in~\cref{sec:Discussion}, we estimate the total meson and muon event rates to demonstrate that these contributions are non-negligible. 

\begin{figure*}
    \includegraphics[width=0.5\textwidth]{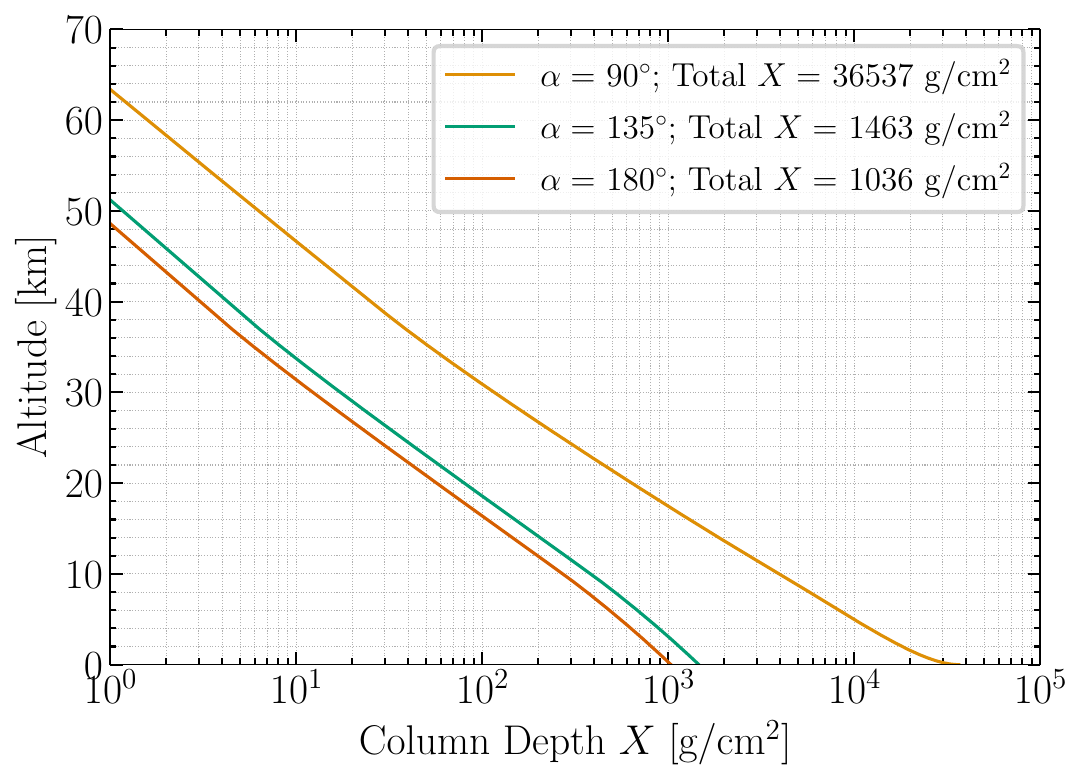}
    \includegraphics[width=0.5\textwidth]{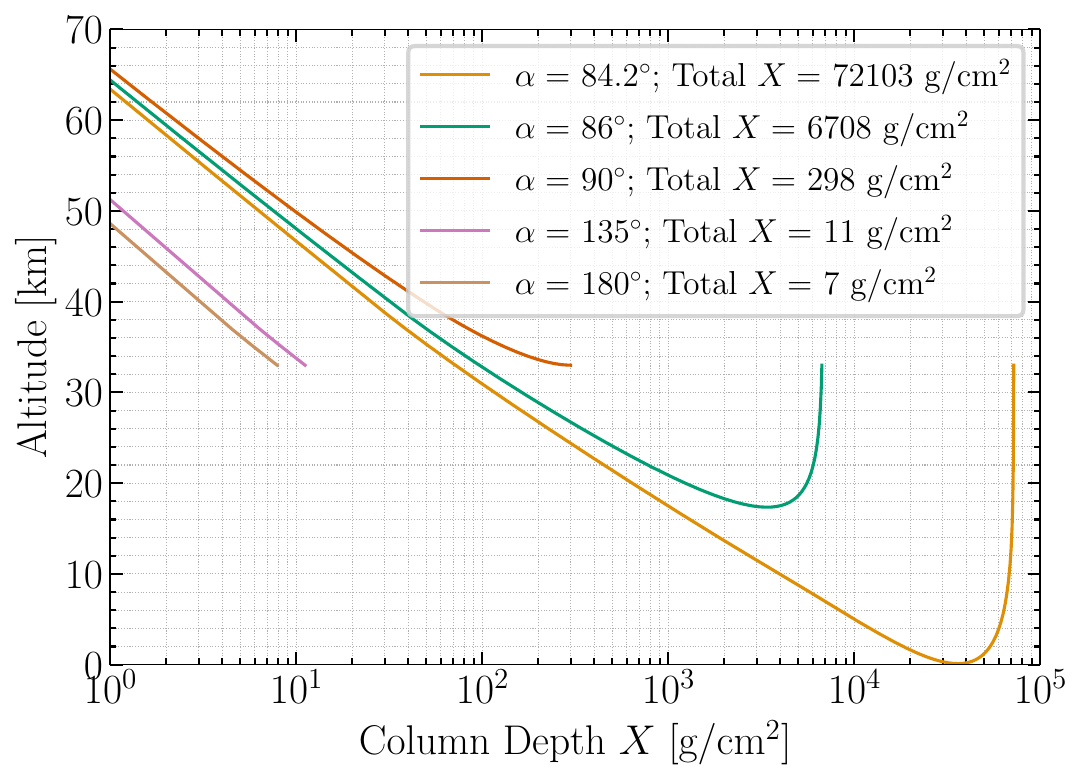}
    \caption{Altitude as a function of atmospheric column depth $X$ for particle trajectories incident from an infinite distance, for trajectories from above and along the detector horizon for $H_0=0$ km (left) and from above, along and below the detector horizon for $H_0= 33$ km (right). The atmospheric density is approximated by the US Standard Atmosphere~\cite{Heck:1998vt}.}
    \label{fig:col_depth}
\end{figure*}

\section{Particle propagation in atmosphere}\label{sec:2}
 We used the open-source Matrix Cascade Equation (MCEq) package~\cite{Fedynitch:2015zma,Fedynitch:2018cbl,Fedynitch:2022vty} to compute the flux of atmospheric muons and mesons ($\pi^{\pm}, K^\pm$) reaching detectors located at various altitudes (0, 18, and 33 km). MCEq numerically solves a set of coupled cascade equations describing the one-dimensional (1-D) propagation and interaction of cosmic ray secondaries through the Earth's atmosphere as a function of column depth $X$. A comprehensive description of the package's implementation and methodology can be found in Refs.~\cite{Fedynitch:2018cbl,Fedynitch:2022vty}. We focus on pion, kaon and muon fluxes for energies larger than 10 GeV where the 1-D approximation is satisfactory~\cite{Lipari:2000du, Lipari:2000wu, Wentz:2003bp, Barr:2004br, Honda:2001xy, Honda:2004yz} (discussed further in~\cref{sec:uncertainties}).

The amount of column depth traversed by particles depends on the trajectory angles $\alpha$, as the Earth's atmosphere is not uniform but decreases in density with increasing altitude. The column depth ($X$) variation with altitude for trajectories with $\alpha=90^\circ,\ 135^\circ$ and $180^\circ$ reaching the surface of the Earth (at 0 km) and sub-orbital altitude (at 33 km) are shown in the left and right panels of~\cref{fig:col_depth}. Also shown in the right panel are altitude-column depth relations for trajectories with $\alpha=84.2^\circ$ and $86^\circ$. For $H_0=33$ km, little atmosphere is above the detector so the column depth is quite small, $X=7$ g/cm$^2$. For the same altitude, a trajectory that just passes the Earth's limb ($\alpha = 84.2^\circ$) traverses a column depth of more than $7.2\times 10^4$ g/cm$^2$, approximately a factor of two times the column depth of a horizontal trajectory that arrives at a detector on Earth ($H_0=0$ km).

The MCEq framework operates by iteratively solving coupled partial differential equations as a function of column depth, discretized into small column depth steps $\Delta X$. At each step, the fluxes of cosmic rays, mesons, and muons are evaluated. The cascade begins at the top of the atmosphere, where only primary cosmic rays are present. These primaries interact with atmospheric nuclei, producing a range of secondary particles such as charged and neutral pions, kaons, protons, neutrons and charm hadrons.
As secondary particles propagate through the atmosphere, their fluxes evolve due to a combination of hadronic interactions and decays. At high energies, interaction processes dominate, shifting the fluxes to lower energies. Below a critical energy threshold ($E_c$), decay becomes the dominant mechanism, further attenuating the flux while producing muons and muon-neutrinos (from pion, kaon and charm hadron decays).

The evolution of the muon flux follows a similar cascade structure. Muons are primarily produced from the decay of charged pions and kaons for $E_\mu \lesssim 10^5$ GeV. Their subsequent propagation also includes electromagnetic energy losses due to ionization and radiative processes. MCEq accounts for these losses to accurately simulate the energy of muons as they travel through the atmosphere and also muon decays which will reduce the muon flux at a given energy that reach the detector.

MCEq provides a range of models for cosmic ray fluxes, atmospheric profiles, and hadronic interaction processes. For our calculations of meson and muon fluxes, we use the Gaisser-Hillas H3a model~\cite{GAISSER2012801} for the primary cosmic ray spectrum, the US Standard Atmosphere~\cite{Heck:1998vt} for the atmospheric profile, and the Sibyll 2.3c model~\cite{Fedynitch:2018cbl} for hadronic interactions. We discuss the impact of alternative cosmic ray flux models and hadronic interaction models in~\cref{sec:uncertainties}. 

\section{Results}\label{sec:results}
In this section, we present the results of various analyses of the charged meson and muon fluxes reaching the detectors. We consider five reference values of 
$\alpha$ to illustrate how the flux varies with the direction of the cosmic ray trajectory: $86^\circ$, $88^\circ$, $90^\circ$, $135^\circ$ and $180^\circ$. The corresponding total column depths for these trajectories, for detectors located at different altitudes, are listed in~\cref{tab:alt_coldepth}. Understanding these column depths is essential for interpreting the flux variation discussed in this section.

\begin{table}[t]
\begin{center}
\begin{tabular}{ |c|c|c|c| } \hline 
 \diagbox[width=8em]{\hspace{11mm}{\boldmath$\alpha$}}{\textbf{Altitude} } & \textbf{0 km} & \textbf{18 km} & \textbf{33 km}  \\ \hline
 86$^\circ$ & - & 55142 g/cm$^2$ & 6708 g/cm$^2$\\ \hline 
 88$^\circ$ & - & 9816 g/cm$^2$ & 938 g/cm$^2$\\ \hline 
 90$^\circ$ & 36537 g/cm$^2$ & 3073 g/cm$^2$ & 298 g/cm$^2$\\ \hline 
 135$^\circ$ & 1463 g/cm$^2$ &  109 g/cm$^2$ & 11 g/cm$^2$ \\ \hline
 180$^\circ$ & 1036 g/cm$^2$ & 77 g/cm$^2$ & 7 g/cm$^2$ \\ 
 \hline
\end{tabular}
\caption{The total column depths in the atmosphere for different trajectory angles $\alpha$ reaching detectors located at different altitudes $H_0$. For $\alpha=86^\circ$ and $88^\circ$, trajectories intersect the Earth when $H_0=0$ km.}
\label{tab:alt_coldepth}
\end{center}
\end{table}

\subsection{Altitude variation}

\begin{figure*}[t]
    \centering
    \includegraphics[width=0.6\textwidth]{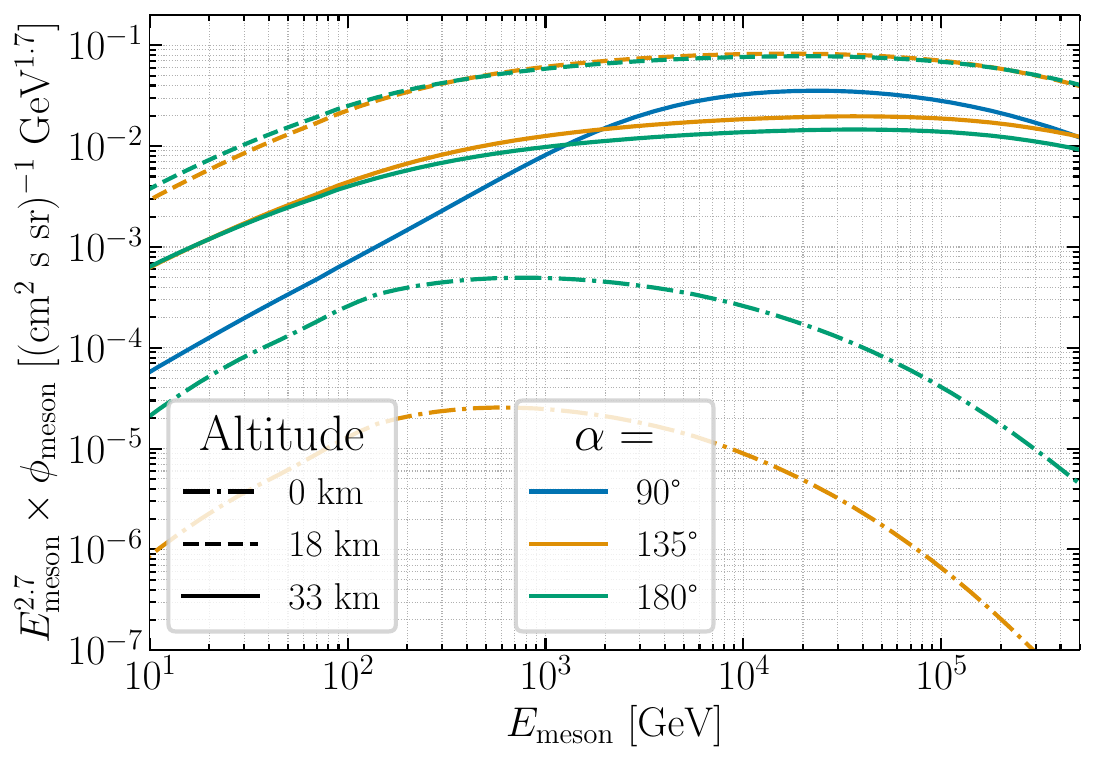}
    \caption{Variation in the  meson ($\pi^{\pm} + K^{\pm}$) flux scaled by $E_{\rm meson}^{2.7}$ as a function of $E_{\rm meson}$ for different detector altitude locations. It is shown for three different $\alpha$ angles. Curves for altitudes $H_0=0,\ 18$ km and $\alpha=90^\circ$ is not shown in the plot because there is a negligible meson flux at the detector. The meson fluxes for $H_0=18$ km and $\alpha=135^\circ$ and $180^\circ$ are nearly equal.}
    \label{fig:meson_altvar_flux}
\end{figure*}

\begin{figure*}[t]
    \centering
    \includegraphics[width=0.45\textwidth]{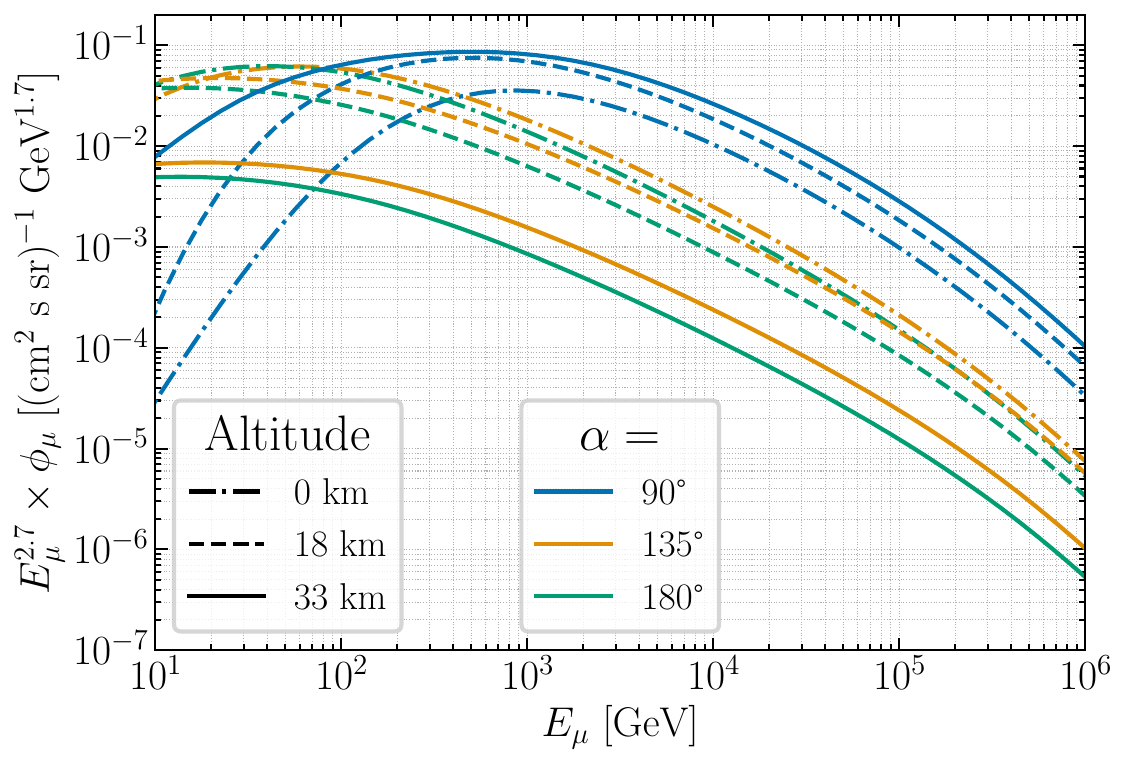}
    \includegraphics[width=0.45\textwidth]{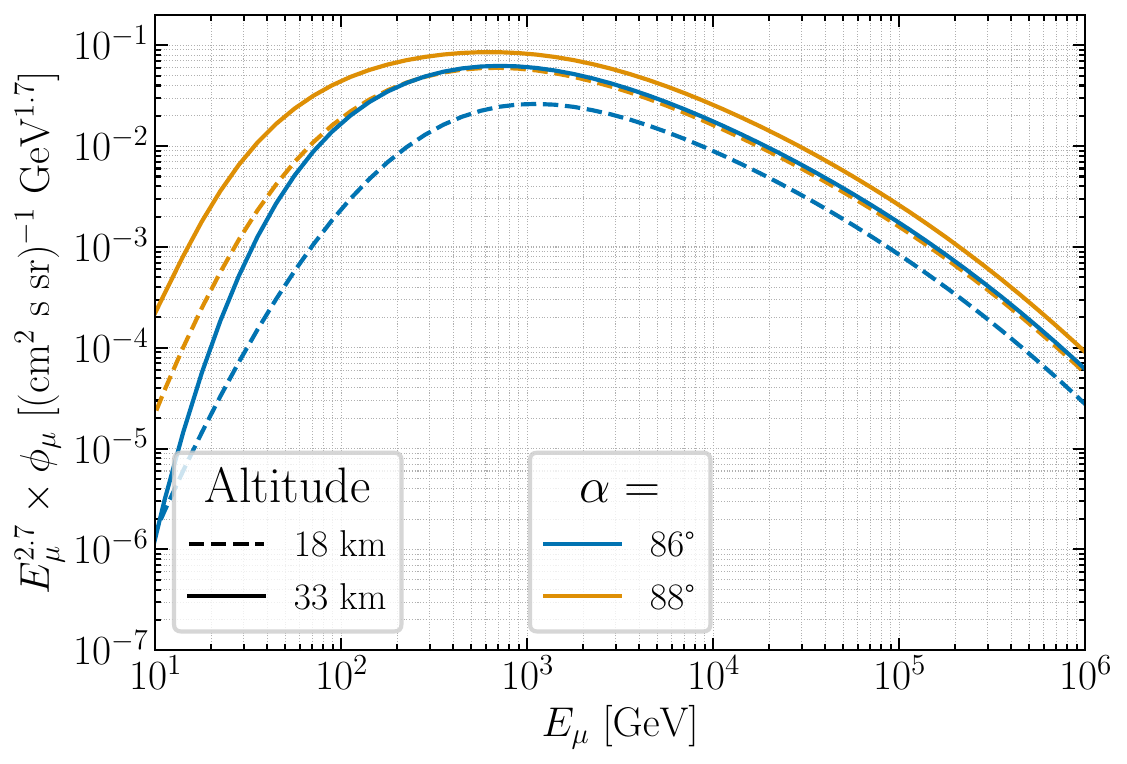}
    \caption{Variation in the  muon flux scaled by $E_\mu^{2.7}$ as a function $E_\mu$ for different detector altitude locations. Left: Three different $\alpha$ angles with horizontal ($\alpha=90^\circ$), downward-going ($\alpha=135^\circ$) and vertical ($\alpha=180^\circ$) trajectories. Right: Upward-going trajectory flux for 18 km and 33 km case. }
    \label{fig:muon_altvar_flux}
\end{figure*}

\Cref{fig:meson_altvar_flux} shows the total flux of pions and kaons, scaled by $E_{\rm meson}^{2.7}$ as a function of the meson energy at the detector, for detectors located at altitudes of 0, 18 and 33 km, and for three different trajectory angles $\alpha$. A general trend is visible across all trajectories: the meson flux increases with energy, reaches a maximum, and then falls off. The energy and angular dependence depends on the relative importance of decays versus interactions. The initial rise is driven by the critical energies of pions and kaons. Below their critical energies, mesons predominantly decay; however, above these critical energies, they are more likely to interact before decaying, leading to an enhanced meson flux. For the vertical flux, the pion critical energy is $E_c ^\pi\approx 110$ GeV and the kaon critical energy is $E_c ^K\approx 850$ GeV~\cite{Gondolo:1995fq}.

At sufficiently high energies, the meson flux scales with energy approximately following the cosmic ray nucleon energy scaling. \Cref{fig:meson_altvar_flux} shows the meson flux scaled by $E_{\rm meson}^{2.7}$. At high energies, the flux of cosmic ray nucleons falls faster than $E^{-2.7}$, so the corresponding energy-scaled meson flux decreases. 

An additional effect is observed when considering different column depths. For low to moderate column depths (i.e., curves for $H_0=18$ km and 33 km), the meson flux has a region of flattening before eventually decreasing at higher meson energies. For higher column depths (i.e., for $H_0=0$ km curves), however, the decrease in meson flux begins earlier and the flattening is less pronounced. This behavior arises when the column depths are considerably larger than the meson interaction length, and leads to  mesons having  more interactions which shift them to lower energies, and reducing the high energy meson component.

The meson flux reaching the detector at $H_0=0$ km is the lowest among all trajectories across the full range of meson energies. This suppression is due to the larger column depth the mesons have to traverse, increasing the probability of interaction (which results in energy loss) or decay before reaching the detector. For $H_0=0$ km and $\alpha=90^\circ$, the meson flux at the detector is highly suppressed because of the large column depth to traverse, so it is not visible on  the scales shown in the figure.

At $H_0=18$ km, the meson flux is higher than at $H_0=33$ km for the $\alpha=135^\circ$ and $\alpha=180^\circ$ trajectories. The longer column depths in these trajectories (for $H_0=18$ km) provide more opportunities for cosmic ray interactions to produce mesons. However, for $\alpha=90^\circ$, the meson flux at $H_0=18$ km is nearly negligible, since mesons encounter particularly large column depths, leading to multiple interactions that decrease their energies before they ultimately decay.

At $H_0=33$ km, the behavior is more nuanced. For low meson energies, the decay probability is higher, so mesons traveling shorter column depths (e.g., along the 
$\alpha=135^\circ$ and $\alpha=180^\circ$ trajectories) are more likely to survive compared to $\alpha=90^\circ$ trajectory. Consequently, the flux is higher for the $\alpha=135^\circ$ and $\alpha=180^\circ$ trajectories in this energy range. However, at higher meson energies, $\alpha=90^\circ$ trajectory has a larger meson flux. This is because the greater column depth for this trajectory provides more opportunities for cosmic ray interactions to produce energetic mesons, which are then sufficiently long-lived to reach the detector. Thus, at high meson energies, the $\alpha=90^\circ$ flux overtakes that of the $\alpha=135^\circ$ and $\alpha=180^\circ$ trajectories.

It is worth noting that the $\alpha=135^\circ$ and $\alpha=180^\circ$ curves are nearly identical for the respective detector altitude $H_0=18$ and 33 km altitude, as the corresponding column depths differ only slightly (as can be seen in~\cref{tab:alt_coldepth}). Additionally, we do not show the meson fluxes for upward-going trajectories in \cref{fig:meson_altvar_flux} because the meson flux arriving at the detector will be negligible as they will mostly decay. 

\Cref{fig:muon_altvar_flux} shows the muon flux scaled by $E_\mu^{2.7}$ at the detector for different altitudes and trajectory angles, the left panel for horizontal or down-going muon trajectories and the right panel for (slightly) up-going muon trajectories for $H_0= 18$ and 33 km. A general trend is observed across all cases: at high muon energies, the atmospheric muon flux falls off roughly as $E^{-(\gamma+1)}$, where $\gamma\simeq 2.7$ is the spectral index of the primary cosmic ray nucleon flux, consistent with the behavior noted in refs.~\cite{Lipari:1993hd,Gondolo:1995fq}. The flattening observed in the curves corresponds to the regime where the muon flux instead follows $E^{-\gamma}$, matching the slope of the primary cosmic ray spectrum.
The suppression seen in some curves at low muon energies is discussed in more detail below, as additional effects play important roles in that regime.

We first look at~\cref{fig:muon_altvar_flux} (left). At $\alpha=90^\circ$, the behavior at low energies is strongly affected by energy losses. At $H_0=0$ km and $H_0=18$ km, nearly all mesons decay before reaching the detector, but the resulting muon flux is still smaller than at $H_0=33$ km. This is because the column depth for the $\alpha=90^\circ$ trajectory is substantial, leading to significant energy losses for muons and causing them to decay. For example, at $H_0=0$ km the average energy loss is $\sim$ 96 GeV over the total column depth, while at $H_0=18$ km it is $\sim$ 8 GeV and for $H_0=33$ km it is $\sim$ 0.7 GeV. This explains the larger separation between the $\alpha=90^\circ$ curves at low muon energies. At higher energies, however, the impact of energy loss becomes less significant.  

The trend for the $\alpha=135^\circ$ and $180^\circ$ curves across all detector altitudes is governed by the requirement of sufficient column depth for cosmic rays to interact and produce mesons, which subsequently decay into muons. The largest flux is observed at $H_0=0$ km, followed by 18 km, and then 33 km -- mirroring the ordering of the corresponding column depths. At lower muon energies, the energy loss for these trajectories is less significant so the strong suppression observed in the $\alpha=90^\circ$ case does not appear here.

~\Cref{fig:muon_altvar_flux} (right) shows the upward-going muon fluxes scaled by $E_\mu^{2.7}$ reaching the detector at $H_0 = 18$ km and $H_0 = 33$ km. In both cases, the behavior at low muon energies is strongly influenced by energy losses. The column depths along each trajectory are large enough that low-energy muons lose a significant fraction of their energy, reducing the energy-scaled flux that reaches the detector. At relatively higher muon energies, however, these losses become negligible compared to the muon’s total energy, and the energy-scaled flux correspondingly increases.

\subsection{Pion/kaon ratios and muon flux fractions}
\begin{figure*}[t]
    \centering
    \includegraphics[width=0.45\textwidth]{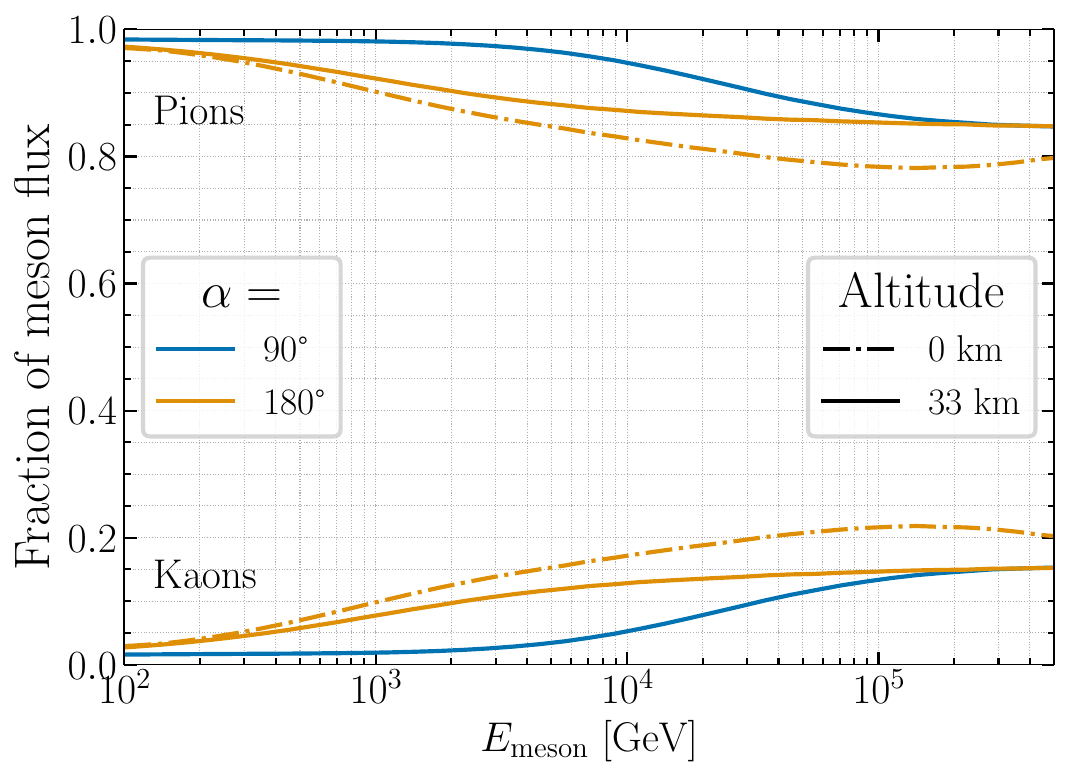}
    \includegraphics[width=0.454\textwidth]{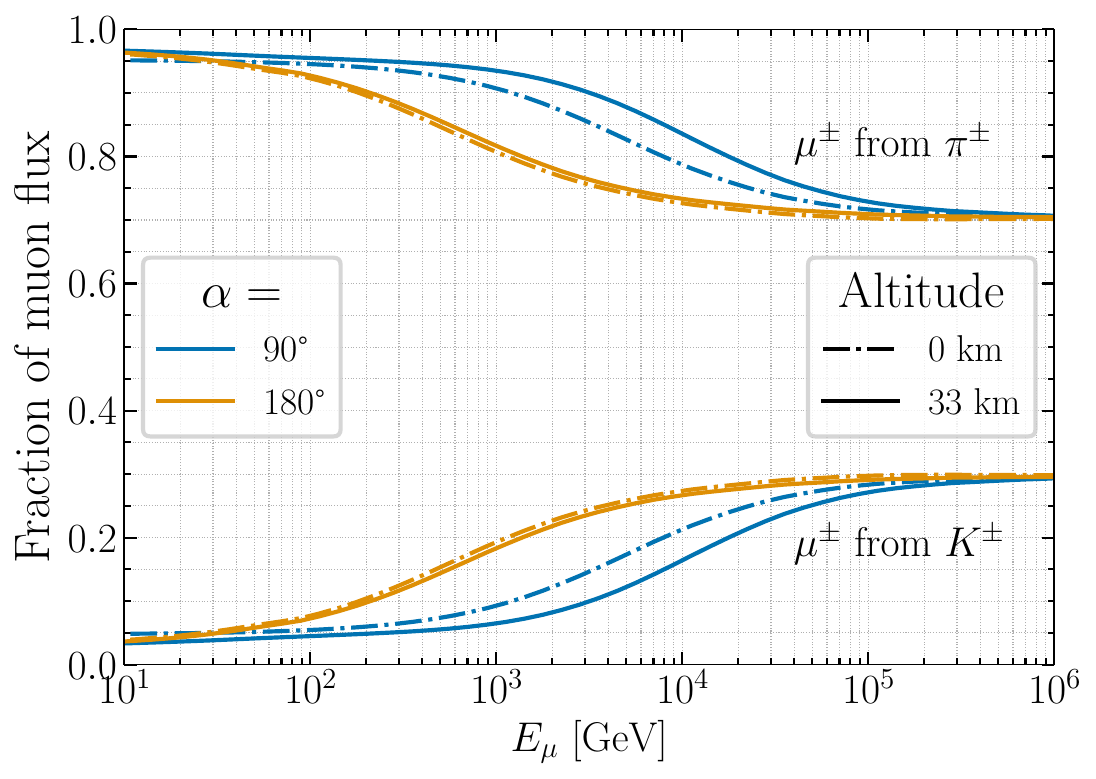}
    \caption{Fraction of mesons (left) and muons (right) reaching detector located at 0 km and 33 km of altitude. The fractions are shown for two trajectories $\alpha=90^{\circ}, 180^\circ$. In the left plot, labels Pions and Kaons represent: $\phi_{CR \to \pi^{\pm}}/(\phi_{CR \to \pi^{\pm}} + \phi_{CR \to K^{\pm}})$ and $\phi_{CR \to K^{\pm}}/(\phi_{CR \to \pi^{\pm}} + \phi_{CR \to K^{\pm}})$, respectively. Curve for altitude $H_0=0$ km and $\alpha=90^\circ$ is not shown because the flux is negligible. For the right plot, the labels represent: $\phi_{\pi^{\pm} \to \mu^{\pm}}/(\phi_{K^{\pm} \to \mu^{\pm}} + \phi_{\pi^{\pm} \to \mu^{\pm}})$ and $\phi_{K^{\pm} \to \mu^{\pm}}/(\phi_{K^{\pm} \to \mu^{\pm}} + \phi_{\pi^{\pm} \to \mu^{\pm}})$. The plots are generated using MCEq package. Note the different $x$-axis scales in the two panels. }
    \label{fig:fraction}
\end{figure*}

~\Cref{fig:fraction} (left) shows the relative contributions of pions and kaons to the charged meson flux as a function of the meson energy ($E_{\rm meson}$) at  detectors at $H_0=0$ km and  33 km and for angles $\alpha=90^\circ$ and $ 180^\circ$. We begin the ratio curves at $E_{\rm meson}=10^2$ GeV, since below this energy essentially all kaons decay and their flux reaching the detectors is negligible. A general trend is observed: pions dominate at lower energies, while the kaon contribution becomes more significant at higher energies, although their relative contributions depend on $\alpha$ and $H_0$. Pions dominate, in part, because of their lower mass, which makes them easier to produce in particle interactions. Additionally, pions are composed of up and down quarks, the same quarks found in protons, whereas kaons contain a strange quark, making their production less favorable.

At low energies, the meson flux ratio remains nearly constant with energy. This initial flattening occurs because both pions and kaons predominantly decay, keeping their relative contributions to the total meson flux nearly unchanged. At high energies, a second flattening occurs, where both meson types interact more often than they decay, again stabilizing their relative contributions. The largest variation appears in the intermediate energy region, where the transition between decay and interaction takes place. Since pions and kaons have different critical energies, the point at which each shifts from decay-dominated to interaction-dominated behavior differs, driving the transition from the first flattening at low energies to the second flattening at high energies.

Differences between curves at various altitudes and trajectories are determined by the effective path length traveled by the particles in the atmosphere. Longer path lengths increase the chances of cosmic ray interactions, leading to the production of additional mesons -- primarily pions. They also allow more decays to occur for both pions and kaons. As a result, the first flattening is extended when the effective path length is larger, as observed for $H_0=33$ km and $\alpha=90^\circ$ curve where the total path length is on the order of several hundreds of km. By contrast, for $\alpha=180^\circ$ trajectories at both $H_0=0$ km and $H_0=33$ km, the path lengths are much shorter, which explains the correspondingly shorter first flattening observed in those curves.

~\Cref{fig:fraction} (right) shows the relative contributions of pions and kaons to the  muon flux for $\alpha=90^\circ$ and $\alpha=180^\circ$ as a function of the muon energy ($E_{\mu}$) at the detector at $H_0=0$ km and $H_0=33$ km. Qualitatively, the muon fractions from pions and kaons follow the meson fractions, however muons come from the mesons that decay rather than from those that survive to reach the detector (shown in the left panel). At low energies, both pions and kaons predominantly decay, leading to the first flattening in the muon flux ratio. As the pion critical energy is reached, their decay probability decreases and their contribution to muons slows, while kaons continue to decay since their much higher critical energy has not yet been reached. This transition region, where pions interact more while kaons still decay, produces the variation in the ratio. At higher energies, both pions and kaons primarily interact rather than decay, resulting in the second flattening, where the relative muon flux from their decays approaches a constant value with energy. 

Overall, pions always contribute more to the muon flux than kaons for two main reasons. First, as noted above, pions are produced more abundantly in cosmic ray interactions. This effect is compounded by the branching ratios, pions decay to $\mu+\nu_\mu$ 99.9\% of the time whereas kaons decay to $\mu+\nu_\mu$ 63.56\% of the time. Additionally,  pion decays give muons most of the pion energy since the muon mass is close to the pion mass. In contrast, kaon decays distribute the available energy more evenly between the muon and the accompanying muon neutrino, producing relatively fewer high energy muons. 

The second flattening in the fractional contributions to the muon flux shown in~\cref{fig:fraction} (right) differs from the results reported in ref.~\cite{Fedynitch:2018cbl} (Figure 29), where a short second flattening is followed by a decrease in the pion and kaon contributions to the muon flux with energy. The difference arises because our analysis considers only the conventional muon flux, namely, muons originating from pion and kaon decays, since we are ultimately interested in the integrated atmospheric flux above 10 GeV. At higher energies ($E_\mu \gtrsim 10^5$ GeV), prompt muons from charm hadron decays become the dominant contribution to the total muon flux.

When comparing vertical ($\alpha=180^\circ$) trajectories at different altitudes, we find that despite the large difference in total atmospheric column depth -- 1036 g/cm$^2$ at 0 km and only 7 g/cm$^2$ at 33 km (from~\cref{tab:alt_coldepth}) -- the pion and kaon contributions to muons remain similar. This apparent similarity arises because the actual path lengths traveled by the cosmic rays and mesons in the atmosphere are not too different. 

The first flattening for the horizontal trajectories ($\alpha=90^\circ$) extends to higher muon energy as compared to the vertical trajectories because of the same reason explained earlier for the meson flux ratios: the longer path lengths for horizontal trajectories. They give more chances for mesons to decay and thus extending the contribution to the muon flux. 

\subsection{Atmospheric flux for different telescope orientations}
\begin{figure*}[t]
    \centering
    \includegraphics[width=0.9\textwidth]{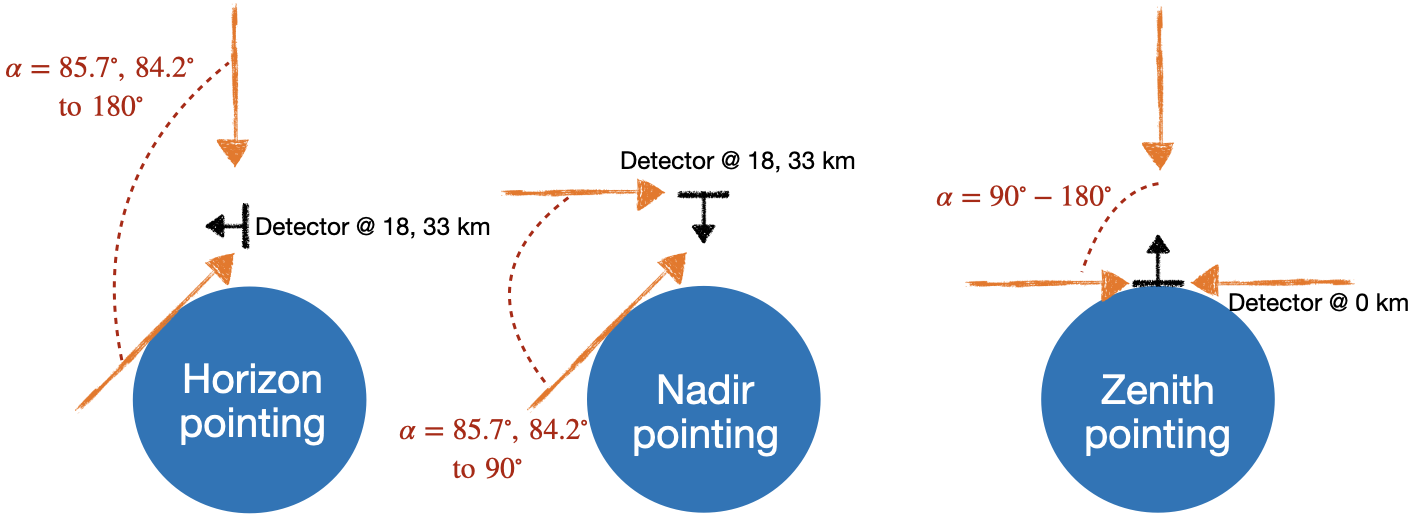}
    \caption{Different orientations of the telescopes located at different altitudes, and the relevant range of $\alpha$ seen by the detectors. For detectors located at $H_0=18$ km and 33 km, the angle made with the Earth's limb are $\alpha=85.7^\circ$ and 84.2$^\circ$, respectively. }
    \label{fig:tel_orientation}
\end{figure*}

\begin{figure*}[t]
    \centering
    \includegraphics[width=0.6\textwidth]{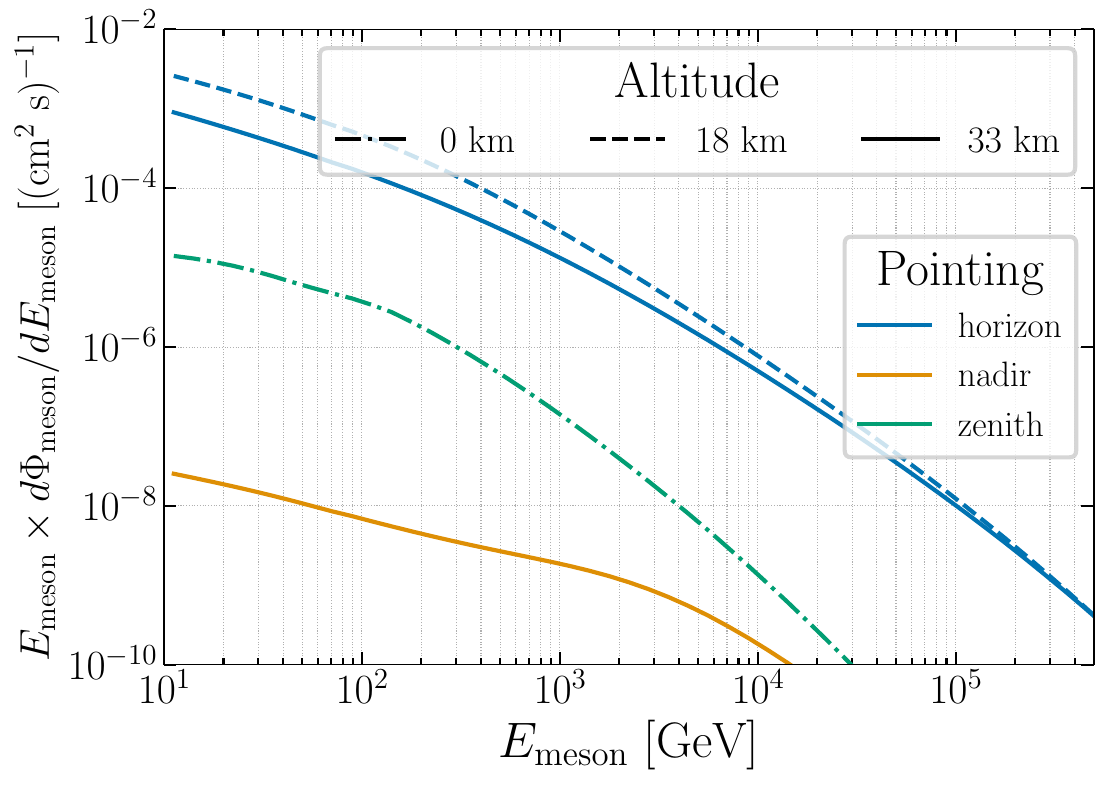}
    \caption{Differential meson flux scaled by meson energy reaching the detector, integrated over the angular range the detector can observe, as a function of the meson energy at the detector. Variations by different detector locations in altitude are plotted. The curve for nadir pointing at 18 km case is not shown because the differential meson flux is negligible.  }
    \label{fig:mesonNum_altvar_flux}
\end{figure*}

\begin{figure*}[t]
    \centering
    \includegraphics[width=0.6\textwidth]{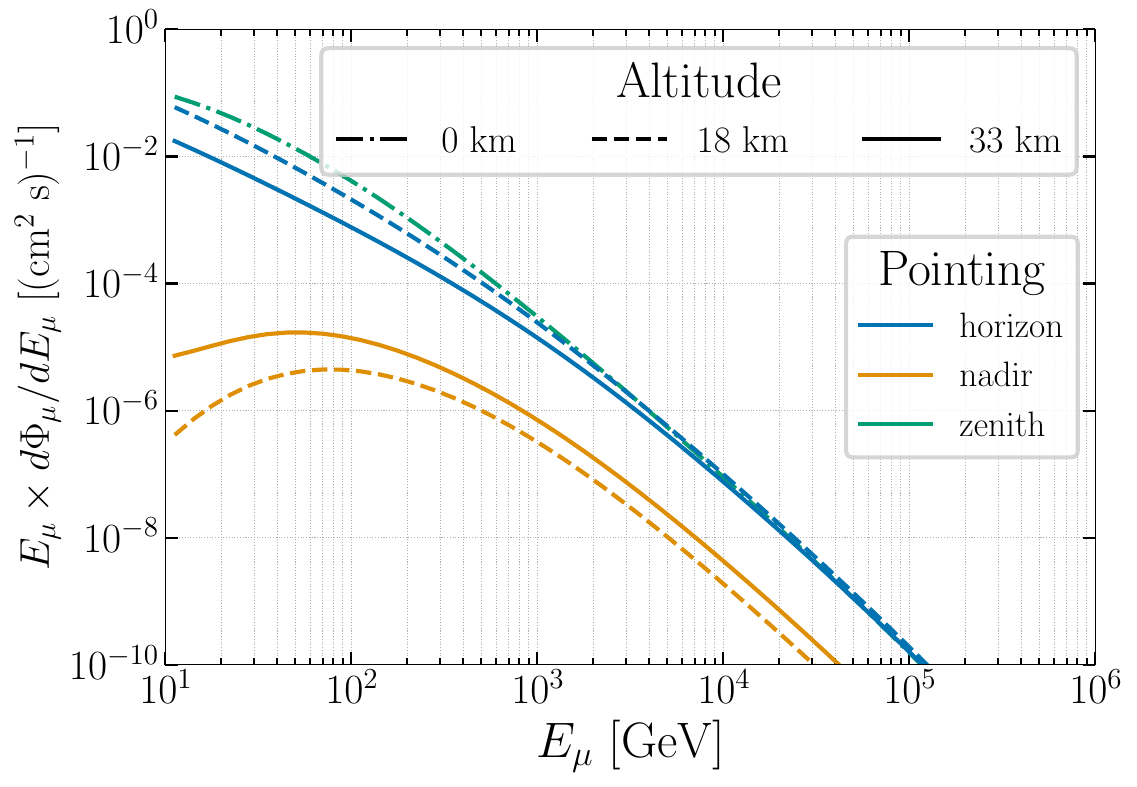}
    \caption{Differential muon flux scaled by muon energy reaching the detector, integrated over the angular range the detector can observe, as a function of muon energy. Variations by different detector locations in altitude are plotted. }
    \label{fig:muNum_altvar_flux}
\end{figure*}

The detectors intercepting meson and muon fluxes can be oriented in different directions when events occur. Here, we consider three pointing orientations: horizon, nadir, and zenith (the latter applying to the 0 km altitude case, where the detector observes the full sky above). Each orientation is sensitive to a distinct range of incoming trajectories, as illustrated in~\cref{fig:tel_orientation}.

To calculate the rate of mesons and muons incident on detectors with these orientations, we integrate the differential flux over the relevant solid angle. The expressions for the differential flux, as a function of  meson or muon energy when it reaches the detector, integrated over solid angle are
\begin{eqnarray}
    \frac{d\Phi_{\mu/\rm{meson}}(E)}{dE} &\equiv& \int_{\varphi=-\pi/2}^{\varphi=\pi/2}\int_{\alpha_{\rm{H, min}}}^{\alpha=180^\circ}\phi_{\mu/\rm{meson}}(E,\alpha) \sin^2\alpha \cos\varphi \, d\alpha 
    d\varphi, \ {\rm horizon} \label{eq:horizontal}\\
    \frac{d\Phi_{\mu/\rm{meson}}(E)}{dE} &\equiv&  \int_{\varphi=0}^{\varphi=2\pi}\int_{\alpha_{\rm{N,min}}}^{\alpha_{\rm{N,max}}}\phi_{\mu/\rm{meson}}(E,\alpha) \cos\alpha \sin\alpha\, d\alpha 
    d\varphi, \ {\rm nadir/zenith}\,. \label{eq:nadir}
\end{eqnarray}
We use both eqs.~(\ref{eq:horizontal}) and (\ref{eq:nadir}) for 18 km and 33 km altitude case with $\alpha_{\rm H, min} = \alpha_{\rm N, min} = 85.7^\circ,\ 84.2^\circ$, respectively, and $\alpha_{\rm N, max} = 90^\circ$. For $H_0=0$ km, we only use eq.~(\ref{eq:nadir}) with $\alpha_{\rm N, min} = 90^\circ$ and $\alpha_{\rm N, max} = 180^\circ$. 

The results for $E_{\rm meson}\times d\Phi_{\rm{meson}}/dE_{\rm meson}$, are shown in~\cref{fig:mesonNum_altvar_flux}. The differential meson flux reaching detectors at 18 km and 33 km altitudes in the horizon-pointing configuration is relatively high. This is primarily due to the broader range of allowed $\alpha$ angles contributing to the differential meson flux. In contrast, for the zenith-pointing detector at 0 km, the differential meson flux is significantly reduced, as mesons traversing these longer column depths tend to decay before reaching the detector.

The nadir-pointing configuration yields the lowest differential meson flux, due to both the narrow angular acceptance and the fact that all contributing trajectories correspond to long column lengths, increasing the likelihood of meson decay. The differential meson flux for the nadir-pointing configuration at 18 km is not shown because it is negligible. 

\Cref{fig:muNum_altvar_flux} shows $E_\mu \times d\Phi_\mu/dE_\mu$.
At lower muon energies, for the zenith- and horizon-pointing configurations, most of the contribution comes from vertical and downward-going angles, with minimal input from more horizontal angles.
As we move to higher muon energies, both upward-going and horizontal angles in the horizon-pointing configuration for 18 km and 33 km cases, and horizontal angles in the zenith-pointing configuration at 0 km case begin to contribute to the differential muon flux, also evident in~\cref{fig:muon_altvar_flux}.

In the case of nadir-pointing, the overall differential muon flux is lower than the other two pointing directions. This is because only trajectories up to $\alpha = 90^\circ$ are included for 18 km and 33 km cases. This limited angular range contributes less to the differential muon flux compared to the broader angular acceptance in zenith- or horizon-pointing configurations.

The total meson and muon fluxes, integrated over energy with a minimum threshold of $E = 10$ GeV, are shown in~\cref{tab:final_flux}. Our use of a 1-D cascade equation approximation for each trajectory angle $\alpha$ is good for energies $E \gtrsim 10$ GeV with some uncertainties which are discussed in~\cref{sec:uncertainties}. 

\begin{table}[t]
\centering
\renewcommand{\arraystretch}{1.5}
\begin{tabular}{|c||c|c|c || c|c|c|}
\hline
\multirow{2}{*}{\textbf{Altitude [km]}} & \multicolumn{3}{c||}{\textbf{Meson Flux [cm$^2$ s]$^{-1}$}} & \multicolumn{3}{c|}{\textbf{Muon Flux [cm$^2$ s]$^{-1}$}} \\ \cline{2-7}
& Horizon & Nadir & Zenith & Horizon & Nadir & Zenith \\ \hline
0   &   -       &   -    &   2.19e-5     &      -      &    -   &  0.0738      \\ \hline
18  &    0.00332      &   4.14e-23    &   -     &   0.0427         &   1.1e-5    &    -    \\ \hline
33  &    0.00114      &  4.34e-8     &  -      &    0.013        &  4.35e-5     &   -     \\ \hline
\end{tabular}
\caption{Meson and muon fluxes for different detector altitudes and pointing directions (direction of normal to the camera surface), integrated over solid angle and final energies greater than 10 GeV. The dashes in the table correspond to pointing orientations that are not geometrically feasible for the given $H_0$, as is also evident in~\cref{fig:tel_orientation}.}
\label{tab:final_flux}
\end{table}

\subsection{Uncertainties}\label{sec:uncertainties}

\input{uncertainties}

\section{Discussion}\label{sec:Discussion}
Over the past decade, there has been a growing effort within the balloon-based astrophysics community to deploy fluorescence and Cherenkov detectors at sub-orbital altitudes for the detection of cosmic ray and neutrino-induced EASs. A key challenge in these missions is accurately characterizing the backgrounds that can mimic or obscure genuine EAS signals. One such background arises from charged particles hitting the detector, particularly mesons and muons, produced in cosmic ray showers. These particles can traverse the detector plane, depositing energy through ionization, and thereby act as a potential background to signals generated by EAS photons.

In this study, we evaluated the flux of charged mesons and muons incident on balloon-borne detectors at altitudes of 18 km and 33 km, and compared these results to a ground-based detector at 0 km. We find that the flux of charged mesons is relatively lower than the flux of muons. In some configurations, the meson flux is about a factor of 10 smaller than the muon flux, while in other configurations, it is much smaller, as seen in \cref{tab:final_flux}. The charged meson flux at detectors may be further reduced by detector shielding.

The atmospheric muon flux remains significant even at balloon altitudes.
To illustrate the scale of the rate of incident muons on a detector, we use the size of the Cherenkov telescope on board the EUSO-SPB2 balloon flight \cite{Adams:2025owi}. With a $16\times 32$ pixel grid of silicon photomultipliers (SiPMs), with each pixel 6 mm $\times$ 6 mm in area, its effective area was 184 cm$^2$. The silicon thickness in each SiPM is 50 $\mu$m. Assuming a 100\%\ detection efficiency for muons with $E_\mu>10$ GeV and using the flux values from \cref{tab:final_flux}, we estimate the atmospheric muon rate. For zenith-pointing at $H_0=0$ km, a detector of this size has an atmospheric muon rate of 13.8 Hz. For a horizon-pointing detector of this size at $H_0=18$ km and 33 km, the atmospheric muon rates are 7.9 Hz and 2.4 Hz, respectively.

The Trinity demonstrator \cite{Bagheri:2024byu,Bagheri:2025fxh}, a ground-based telescope with a Cherenkov camera half the size of the EUSO-SPB2 Cherenkov camera, located on Frisco Peak in Utah, USA  has recorded tracks that may possibly come from muons \cite{trinity-icrc}. Their track rate is approximately one per hour. For the tracks to be interpreted as muons, the muons would traverse the silicon through several pixels, entering a SiPM through the 6 mm $\times$ 50 $\mu$m edge (the “silicon edge’’) rather than through the 6 mm $\times$ 6 mm surface. For a planar Cherenkov camera oriented toward the horizon, the silicon edge of the camera is zenith-pointing. The relevant flux would be a modification of the zenith-pointing flux, adjusted to accommodate a limited solid angle. Assuming that at least 4 pixels are triggered per “track-like” event, incident muons that generate these track-like events range in $\alpha = 166^\circ-180^\circ$, the flux for $E_\mu>10$ GeV is $5.56\times 10^{-3}$ cm$^{-2}$s$^{-1}$ for $H_0=0$ km. For a single pixel surface area of 6 mm $\times$ 50 $\mu$m, the rate at $H_0=0$ is 0.06 muons/hr. The Trinity camera is 16 pixels wide, so across the silicon edge of the whole camera, the muon rate to make 4 pixel-length (or more) tracks with these approximations is 0.96 muons per hour for $E_\mu>10$ GeV. If such a camera were pointed up (zenith-pointing) where 92 cm$^2$ surface area can intercept muons, the rate for $E_\mu>10$ GeV is 6.8 Hz. 

A full accounting of the rate of incident atmospheric muons that can be detected requires a more detailed study of detector positioning and altitude. For example, the Trinity demonstrator is at $H_0=3$ km. Additionally, detector-specific analysis must be made to determine whether or not muons can be detected. 
The actual energy a muon deposits in the SiPM described above depends on the thickness of the silicon traversed. For the zenith-pointing case, muons pass through only $50\ \mu\text{m}$ of active material. Using the minimum ionization energy loss for muons in silicon, $\langle dE/dX\rangle \simeq 1.664\ \text{MeV\,cm}^2/$g~\cite{Groom:2001kq}, a muon deposits $\sim 19.4$ keV of energy. By contrast, in the earlier case where muons crossed the thinner silicon edge and passes lengthwise through a pixel ($6$ mm path length), the expected energy deposit is $\sim 2.33$ MeV. Whether such energy deposits are detectable, and whether or not they are recorded, is a crucial experimental consideration. For example, EUSO-SPB2's Cherenkov telescope had bifocal optics. Triggered events required two (bi-focal) pixels to reach a set threshold in a 10 ns time window \cite{Adams:2025owi}. In addition, as discussed in~\cref{sec:uncertainties}, different hadronic interaction models lead to variations of up to $\sim 5-20\%$ in the predicted muon flux. Finally, the muon fluxes in \cref{tab:final_flux} are integrated with $E_\mu>10$ GeV. At lower muon energies, location-dependent 3-D effects in the cascade equations become important, as noted in~\cref{sec:uncertainties}. A 3-D atmospheric propagation analysis is beyond the scope of the present study.

\vspace{0.2cm}
\section*{Acknowledgements}
We thank J. Krizmanic, J. Szabelski, and D. Fuehne for illuminating discussions, C. Feltman for helpful insights on the effects of the Earth’s magnetic field, and S. Stepanoff for discussions regarding the Trinity experiment. This work was supported in part by the U.S. Department of Energy under grant DE-SC-0010113.

\bibliographystyle{elsarticle-num}
\bibliography{references}

\end{document}

%% file: units.tex
\DeclareSIUnit \s {\second}
\DeclareSIUnit \ns {\nano\second}
\DeclareSIUnit \mus {\micro\second}
\DeclareSIUnit \ms {\milli\second}
\DeclareSIUnit \MB {\mega\byte}
\DeclareSIUnit \GB {\giga\byte}
\DeclareSIUnit \TB {\tera\byte}
\DeclareSIUnit \PB {\peta\byte}
\DeclareSIUnit \Mbps {\mega\bit/\s}
\DeclareSIUnit \Gbps {\giga\bit/\s}
\DeclareSIUnit \Tbps {\tera\bit/\s}
\DeclareSIUnit \Pbps {\peta\bit/\s}
\DeclareSIUnit \kton {\kilo\tonne} 
\DeclareSIUnit \kt {\kilo\tonne}
\DeclareSIUnit \Mt {\mega\tonne}
\DeclareSIUnit \eV {\electronvolt}
\DeclareSIUnit \keV {\kilo\electronvolt}
\DeclareSIUnit \MeV {\mega\electronvolt}
\DeclareSIUnit \GeV {\giga\electronvolt}
\DeclareSIUnit \TeV {\tera\electronvolt}
\DeclareSIUnit \PeV {\peta\electronvolt}
\DeclareSIUnit \EeV {\exa\electronvolt}
\DeclareSIUnit \m {\meter}
\DeclareSIUnit \cm {\centi\meter}
\DeclareSIUnit \in {\inchcommand}
\DeclareSIUnit \km {\kilo\meter}
\DeclareSIUnit \kV {\kilo\volt}
\DeclareSIUnit \kW {\kilo\watt}
\DeclareSIUnit \MW {\mega\watt}
\DeclareSIUnit \MHz {\mega\hertz}
\DeclareSIUnit \mrad {\milli\radian}
\DeclareSIUnit \year {years}
\DeclareSIUnit \POT {POT}
\DeclareSIUnit \sig {$\sigma$}
\DeclareSIUnit\parsec{pc}
\DeclareSIUnit\lightyear{ly}
\DeclareSIUnit\foot{ft}
\DeclareSIUnit\ft{ft}
\DeclareSIUnit \ppb{ppb}
\DeclareSIUnit \ppt{ppt}
\DeclareSIUnit \samples{S}
\DeclareSIUnit \pe{PE}
\DeclareSIUnit \T{T}

\newcommand{\enu}{\E_\enu}

%% file: uncertainties.tex
In our 1-D evaluation of the atmospheric muon flux, we have taken $E_\mu>10$ GeV to mitigate corrections from the Earth's magnetic field and geometric effects that require a 3-dimensional (3-D) treatment. Uncertainties in our integrated atmospheric flux rates come from uncertainties in the incident cosmic ray flux, hadronic interactions, and 3-D effects.

The dominant contribution to the muon flux in our rate calculations comes from muons with energies up to $\sim 100$ GeV. The relevant primary cosmic ray energies producing these muons range from 10's of GeV to $10^4$ GeV. The flux of low energy cosmic rays is affected by solar activity~\cite{Gleeson:1968zza, Potgieter:2013pdj}, confirmed by the BESS and PAMELA experiments~\cite{Abe:2015mga, Marcelli:2020uqv}. Measurements show that solar modulation is negligible for cosmic ray protons above $\sim 30$ GeV and for helium nuclei above $\sim 15$ GeV per nucleon, so solar modulation has a negligible impact on the resulting fluxes of muons and mesons with energies above 10 GeV.

At higher energies, the cosmic ray spectrum and composition may impact the muon and meson fluxes at detectors.
As discussed in ref.~\cite{Lipari:2019jmk} (and references therein), changes in the cosmic ray spectral index have been measured at $\sim 500$ GeV and $\sim 10^4$ GeV. To test their impact, we implemented a modified broken power-law spectrum with $\gamma=2.67$, 2.85, and 3 at 500 GeV, $10^4$ GeV, and $10^6$ GeV, respectively. We found that these changes in the cosmic ray spectrum had a negligible effect on the resulting muon flux for $E_\mu>10$ GeV for the range of trajectory directions discussed here. A more comprehensive examination of the impact of cosmic ray spectrum and composition uncertainties on the atmospheric lepton fluxes appears in ref. \cite{Yanez:2023lsy}, where they show that the majority of the uncertainties appear at cosmic ray nucleon energies above $\sim 10^4$ GeV. These uncertainties do not feed down significantly to $E_\mu\sim 10-100$ GeV.

Uncertainties in hadronic interactions, especially in the secondary particle yields in cosmic ray nucleon-air interactions, introduce uncertainties in the atmospheric lepton fluxes. 
In this paper, we have focused on muon energies of $E_\mu \gtrsim 10$ GeV. 
A comparison of the vertical muon flux at sea level using different hadronic interaction models indicates deviations of at most $\sim 10\%$ relative to Sibyll~2.3c model \cite{Fedynitch:2018cbl} for  $E_\mu=10–100$ GeV. Other comparisons show $\sim 5-20\%$ uncertainties in the muon flux at sea level associated with the interaction models \cite{Fedynitch:2022vty}. 

Seasonal temperature variations change the atmospheric density profile, thereby changing the altitudes of first and subsequent interactions of incident cosmic rays and the hadrons they produce. This feeds down to the atmospheric flux of muons and neutrinos. The IceCube collaboration has measured a $3.9-4.6\%$ effect for lepton energies in the range $125~\text{GeV}-10~\text{TeV}$ ~\cite{IceCube:2025wtk}. We expect a similar effect for the meson fluxes. 
    
The final topic on uncertainties in our evaluation of muon and meson rates at detectors is tied to the 1-D approximation we use to evaluate the atmospheric fluxes for energies above 10 GeV. In the 1-D approximation, we neglect magnetic field effects ~\cite{Lipari:1998cm, Lipari:2000wu, Wentz:2003bp, Barr:2004br, Honda:2001xy, Honda:2004yz,Hansen:2004kf} and geometrical effects that depend on zenith angle when the meson and lepton production is not collinear with the incident cosmic ray trajectory ~\cite{Lipari:2000wu,Battistoni:1999at}. 
Both effects are more important at lower cosmic ray energies than at higher energies. To account for these, several groups have implemented 3-D simulations of cosmic ray interactions to calculate the fluxes of muon neutrinos and electron neutrinos at the Earth’s surface~\cite{Barr:2004br, Honda:2001xy, Honda:2004yz}. The largest differences between 1-D and 3-D results come from azimuthal variations that produce the well-known East-West asymmetry~\cite{Lipari:2000du}. 
However, since our work integrates over the full azimuthal range, these differences effectively cancel out. 

There are 3-D corrections to the fluxes at different zenith angle (different $\alpha$) as well.
For muon neutrinos in the $1-10$ GeV range, neutrinos originate primarily from cosmic rays with energies of about $3-300$ GeV. The same range of cosmic ray energies  corresponds to muons with energies of $1–$ few tens of GeV.  
The 3-D sea level neutrino flux simulations  differ from 1-D neutrino flux results by $\sim 10-15\%$ for $E_\nu \sim 1-3$ GeV, in near-horizontal directions. For near-vertical directions, the difference is much smaller. 
For $E_{\nu_{\mu}} \gtrsim 10$ GeV, the results of 1-D and 3-D calculations converge (see figs.~9 and 10 in ref.~\cite{Honda:2004yz}). 
Overall,
we expect that the variation in the integrated atmospheric muon flux for $E_\mu>10$ GeV due to geomagnetic effects on cosmic rays and muons, and geometrical effects, is  at most $\sim 10-15\%$ at sea level. Similar uncertainties are expected for fluxes at balloon altitudes ($H_0=33$ km), since $H_0$ is small on the scale of the radius of the Earth.